\documentclass[graybox]{svmult}

\usepackage{mathptmx}       
\usepackage{helvet}         
\usepackage{courier}        
\usepackage{type1cm}        
\usepackage{makeidx}         
\usepackage{graphicx}        
\usepackage{multicol}        
\usepackage[bottom]{footmisc}
\usepackage{tikz}
\usepackage{xcolor}
\usetikzlibrary{matrix,shapes,arrows,positioning,chains}
\usepackage{hyperref}

\makeindex             

\begin{document}
\title*{Detecting and classifying moments in basketball matches using sensor tracked data}
\author{Tullio Facchinetti and Rodolfo Metulini and  Paola Zuccolotto
	}
\institute{
Tullio Facchinetti \at Department of Industrial, Computer and Biomedical Engineering, University of Pavia, Via Ferrata, 1, 27100 Pavia, \email{tullio.facchinetti@unipv.it}
\and Rodolfo Metulini \at Department of Economic and Management, University of Brescia, Contrada Santa Chiara, 50,
	25122 Brescia, \email{rodolfo.metulini@unibs.it}
\and Paola Zuccolotto \at Department of Economic and Management, University of Brescia, Contrada Santa Chiara, 50, 25122 Brescia, \email{paola.zuccolotto@unibs.it}
\and \textbf{Please cite as:} Facchinetti, T., Metulini, R., Zuccolotto, P. (2019) Detecting and classifying moments in basketball matches using sensor tracked data. SIS 2019 - Smart Statistics for Smart Applications - Book of short papers, editors: Giuseppe Arbia, Stefano Peluso, Alessia Pini, Giulia Rivellini. ISBN 9788891915108.
}
%
%
\maketitle

\abstract{Data analytics in sports is crucial to evaluate the performance of single players and the whole team. The literature proposes a number of tools for both offence and defence scenarios. Data coming from tracking location of players, in this respect, may be used to enrich the amount of useful information. In basketball, however, actions are interleaved with inactive periods. This paper describes a methodological approach to automatically identify active periods during a game and to classify them as offensive or defensive. The method is based on the application of thresholds to players kinematic parameters, whose values undergo a tuning strategy similar to Receiver Operating Characteristic curves, using a ground truth extracted from the video of the games.}

\section{Introduction}
\label{sec:intro}

Recent years registered the rise of data analytics applied to sports.
Experts in data science are nowadays employed by teams to improve strategical decisions. 
These strategies, related to the objective of winning the games, may regards either single players or the whole team and must consider both offensive and defensive performances.
In basketball, Oliver~\cite{oliver04rules} outlines most of the tools used to evaluate performance.
Offensive performance can be measured in terms of shots (Zuccolotto et al.~\cite{zucco18big}) or considering other aspects of playing, such as the number of possessions per game (Kubatko et al.~\cite{kubatko07starting}).
Sampaio et al.~\cite{sampaio15exploring} and Paulauskas et al.~\cite{paulauskas18basket} applied a descriptive discriminant analysis to different competitions to identify which variables best predict differences in the playing style.
From a defensive perspective, Franks et al.~\cite{franks15counter} and Goldsberry and Weiss~\cite{goldsberry15effect} introduced a new suite of defensive metrics suggesting to integrate spatial approaches and player tracking data.

The positioning and the velocity of players is an essential aspect to be considered when analysing both offensive and defensive performance. 
The robust statistical apparatus of National Basketball Association (NBA) supported by private companies made analysis of offensive and defensive moments relatively easy.
Wu and Bornn~\cite{wu2017modeling} provide a tool for the visual analysis of offensive actions using a sensor data technology. Miller and Bornn~\cite{miller2017possession} use the same data to catalogue NBA league strategies according to players' movements.
Ball circulation during offense actions has been analysed by D'Amour et al.~\cite{d2015move} to show that the more open shots opportunities can be generated with more frequent and faster movements of the ball. 
Less attention was paid to European leagues, mainly depending on the restriction on data collection, which is rarely granted to authorized operators.
Metulini et al.~\cite{metu17space,metu17sensor} and Metulini~\cite{metu18mining} use tracked data from Italian professional basketball games collected by mean of an accelerometer device and they split games into clusters of homogeneous spatial distances among players, looking for those with better team shooting performance.
Metulini et al.~\cite{metu18area}, using the same data, measured the relation of surface area occupied by players in offence and in defence with the number of scored points by the team.  

However, accelerometer devices track players' movements along the full game, without distinguishing between active/inactive periods and offensive/defensive possessions.
A possible solution is to instruct a person to track these information during the game. 
However, this option can be unpractical either due to organizational issues and cost impact.
For these reasons, we propose a procedure that identifies and removes inactive periods and classifies them as either offensive or defensive.
Such a procedure allows a better usage of tracked data from localization systems for the aim of a better team performance analysis at support to game decision in basketball, from professional to amateur and youth leagues. 

In this paper we discuss the procedure proposed by Metulini~\cite{metu17filt} and we introduce a validation strategy based on the use of a ``ground truth'' extracted from a video-based annotation of a sample of games.
\section{Data}
\label{sec:data}
The tracking system collects the position and the velocity of every player during the full game length, including whose waiting on the bench, along the $x$-axis (court length) and the $y$-axis (court width).
The measured positions are expressed in centimeters (cm); the estimated accuracy of the tracking system is around $30$ cm.
Each measurement is marked by its time instant $t$.
The tracking system is able to capture measurements at a sampling frequency of $50$ Hz, corresponding to a measurement every $20$ milliseconds (ms).
We call the ordered set of measurements \textbf{X}.
The measurement made at the time instant $t$, denoted with $x_t$, thus contains the following information:
\begin{itemize}
	\item The vector of the position for the $i$-th player along the $x-$ and the $y-$ axis, denoted as $pos_i(t) = \{pos_{i}^x(t), pos_{i}^y(t)\}$, measured in cm, where superscript $x$ and $y$ are used, respectively, for court length and court width;
	\item The vector of the velocity for the $i$-th player along the $x-$ and the $y-$ axis, denoted as $vel_i(t) = \{vel_{i}^x(t), vel_{i}^y(t)\}$, measured in kilometres per hour (km/h);
	\item The velocity for the $i$-th player in the court at time $t$, computed as $v_i(t) = \sqrt{vel_i^x(t)^2 + vel_i^y(t)^2}$.
\end{itemize}
\section{The procedure}
\label{sec:proc}
\begin{figure}[htbp!]
	\centering
	\includegraphics[width=\linewidth]{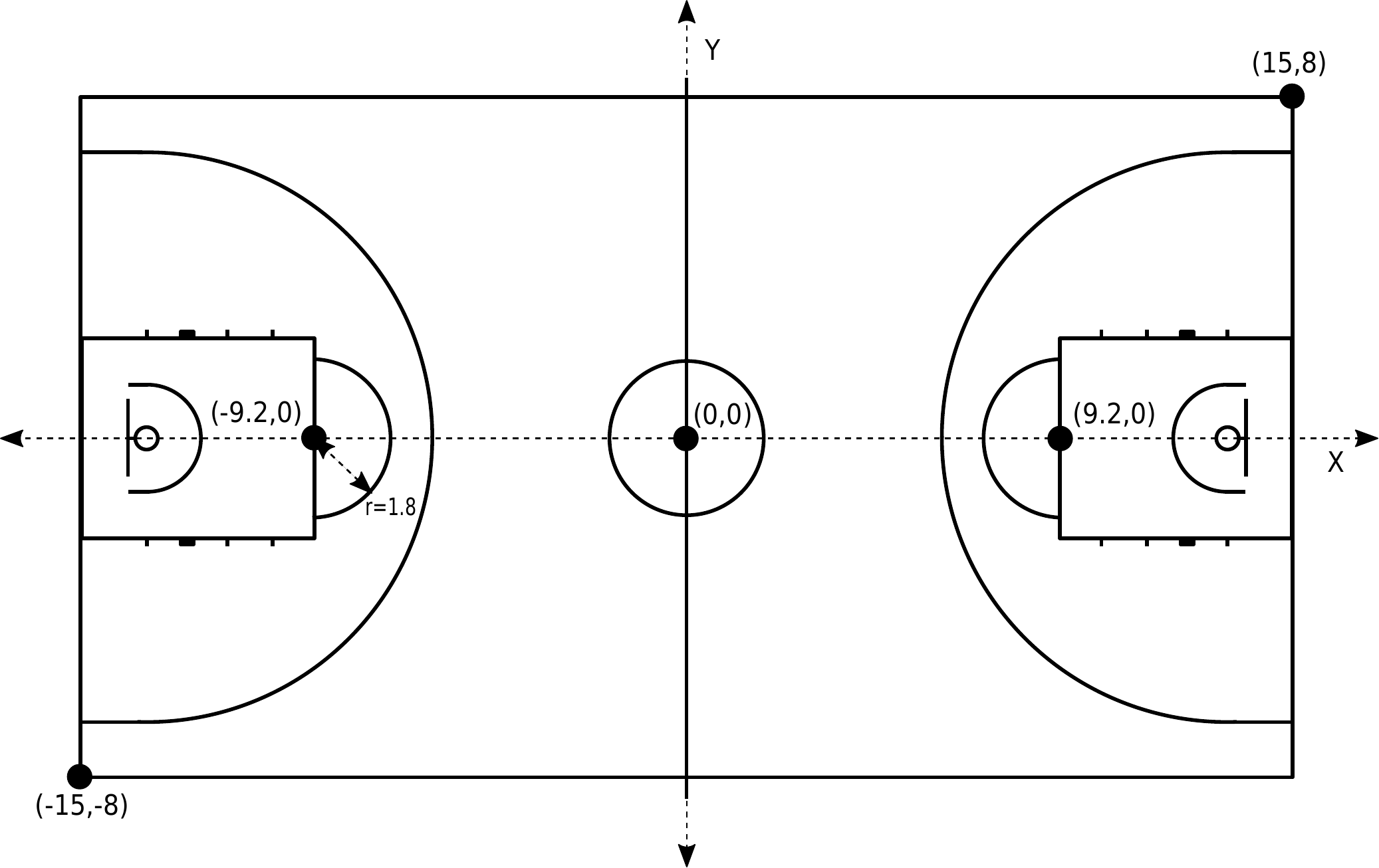}
	\caption{International Basketball Federation (FIBA) court with relevant measures annotated.}
	\label{fig:court_fiba}
\end{figure}

The procedure aims at removing specific measurements from $\mathbf{X}$ according to three different criteria and to separate the game into offensive and defensive possessions. 
The filtering and labelling scheme is based on defining kinematic parameters related to players' positions and velocities on the International Basketball Federation (FIBA) court (Figure~\ref{fig:court_fiba}).
The outcome is a reduced set of measurements denoted as $\mathbf{Xr}$ that includes two features about the type of possession ($poss$ = \{offensive, defensive, transition\}) and the ordered number of possession ($ord$ = \{1,2, ..., n\}), respectively. 
In detail:
\begin{enumerate}
	\item According to criterion 1-A, the procedure drops from \textbf{X} all the measurements belonging to the time instant $t$ in which the number of players inside the court is different from $5$.
	\item Criterion 1-B drops from \textbf{X} the measurements when at least one  player is on the free throw shooting area (FTSA) for at least a specified interval of time $T_{ft}$.
	Player $i$ lies in the FTSA at time $t$ if the vector $pos_i(t)$ lies in the circle $C_r$ centred on the center of the FTSA of radius $r= 1.80m$.
	\item The criterion 1-C removes those measurements where the speed of all the five players in the court is below a given threshold $V_{min}$, for an interval of time equal or larger than $T_{vel}$.	
    \item Criterion 2-A assigns the value of the variable $poss$ to each measurement of $\mathbf{Xr}$. 
    The procedure generate the average $x$ coordinate of the five players on the court at time $t$, $avg\_pos^x(t) = \sum_{i=1}^{5}pos_{i}^x(t) / 5$.
    The measurement $t$ could lie either on the offensive ($avg\_pos^x(t) > 4$) or on the defensive ($avg\_pos^x(t) < -4$) side of the court. \textit{transition} is instead assigned to variable $poss$ whereas $avg\_pos^x(t)$ lies in the interval \{+4,-4\}. 
	\item Criterion 2-B assigns to the measurement of $\mathbf{Xr}$ the $ord$ value, for the aim of counting for the total number of possessions in the game. 
	The procedure assign to $ord$ value ``1'' to measurement $x_1$. 
	Such a value increases by 1 whenever the variable $poss$ takes the value $transition$ at time $t-1$ and either $offensive$ or $defensive$ at time $t$.
\end{enumerate}
\section{The validation strategy}
\label{sec:strat}
While parameter $T_{ft}$ can easily be determined by looking to the average time required by a player to shot one or two free-throws, the ``best'' values for the parameters  $V_{min}$ and $T_{vel}$ need for a tuning strategy in order to be correctly identified.
The tuning strategy has the objective to find those values for $V_{min}$ and $T_{vel}$ such that the accordance of the procedure with the ``ground truth'' is maximized. 
\subsection{Video-based annotation}
\label{sec:vban}
We extract the ``ground truth'' by using a specific smartphone application while watching the available video footage of the game. 
We take note of a number of game events related to i) the moments in which the game is active/inactive; ii) the moments of free-throws, time-outs, quarter- and half-time intervals; iii) the moments when the team was in offence/defence (Table~\ref{buttons}). 
Based on the recorded data, we produce two reports.
\begin{table}[htbp!]
	\sidecaption
	\begin{tabular}{lll}
		\textbf{game event} && \textbf{description} \\
		\hline
		start free-throws && a player is on the FTSA to shoot a free-throw \\ 
		stop free throws && the game stops due a free-throw \\
		start time-out && a time-out starts\\
		stop time-out  &&  a time-out ends \\
		start half-time interval && an half-time interval starts  \\
		stop half-time interval &&  an half-time interval ends \\
		start quarter-time interval && a quarter-time interval starts  \\
		start quarter-time interval && a quarter-time interval ends \\
		stop  &&  the game stops for a generic reason \\
		start &&  the game starts / restarts after a generic stop  \\
		offence &&  the team starts an offensive action \\
		defence && the team starts a defensive action \\	
		\hline	
	\end{tabular}	
	\caption{Names of the events and description.\label{buttons}}
\end{table}
The first report displays when the action starts to be active (\textit{action = play}) or starts to be inactive (\textit{action = stop}) with reference to a given moment (\textit{sec}). 
\texttt{active} is a variable that assumes value equal to $1$ if the game starts to be inactive in that moment. 
\texttt{timeout}, \texttt{ft}, \texttt{quarter} and \texttt{half} are variables assuming value equal to $1$ if the reason of the inactivity is, respectively, a time-out, the shooting of a free-throw, a quarter-time interval or an half-time interval. 
In the excerpt reported in Table~\ref{tab:rep1}, the game starts at second $1$ ($active=0$ \& $sec=1$ in the first row). From second $1$ to second $4$ the game is active. 
At second $5$ the game stops ($active=1$ at the second row of the table) due to a generic reason. 
At second $13$ the game restarts (third row) and at second $47$ the game stops due to a free-throw ($ft = 1$ in the fourth row).
The second one reports the variable \textit{action}, which can assumes either value 1 (start a offensive action for the team) or value 0 (start a defensive action for the team). 
In the excerpt in Table \ref{tab:rep2} the team starts the game in defence ($sec$ = 1), it starts an offensive play at second 32,  it goes back to defence at second 72, an so on.
\begin{table}[htbp!]
	\parbox{.45\linewidth}{
		\sidecaption
		\begin{tabular}{ccccccc}
			\textbf{action}& \textbf{sec}	 &\textbf{active} &	\textbf{timeout}& \textbf{ft}	& \textbf{quarter} &	\textbf{half} \\
			\hline
			play	&1	&1	&0	&0	&0	&0\\
			stop	&5&	0	&0&	0	&0&	0\\
			play	&13	&1	&0	&0&	0	&0\\
			stop	&47&	0&	0	&1	&0	&0\\
			\hline	
		\end{tabular}	
		\caption{An excerpt from the first report.}
		\label{tab:rep1}
	}
	\hfill
	\parbox{.45\linewidth}{
		\begin{tabular}{ccc}
			\textbf{action}& 	\textbf{sec} & \textbf{off}\\
			\hline  
			off&	1& 1\\
			def&	32& 0\\
			off&	72& 1 \\
			def&	138& 0 \\
			\hline	
		\end{tabular}	
		\caption{An excerpt from the second report.}
		\label{tab:rep2}
	}
\end{table}
\subsection{The ``ROC'' method}
\label{sec:appr}
We use the ground truth to check for the robustness of the classification of the procedure in relation to the choice of parameters $V_{min}$ and $T_{vel}$.
Actually, the robustness may be evaluated either in terms of how the procedure classifies active/inactive moments or in terms of how it classifies offence and defence.
We borrow the approach of the Receiver Operating Characteristic (ROC) curves.
The Area Under the Curve (AUC) is traditionally computed based on sensitivities and specificities to quantify the robustness of a prediction method (Zhou et al.~\cite{zhou2009statistical}, Pepe~\cite{pepe2003statistical}, Krzanowski and Hand~\cite{krzanowski2009roc}). 
Sensitivity measures the proportion of \textit{true} positives, while specificity measures the proportion of \textit{true} negatives.
The ROC and AUC help in deciding the optimal threshold value by computing sensitivity and specificity on a series of possible thresholds.
In this problem we have no threshold values to set for the underlying probabilities.
In our case, the measurements are directly classified by the procedure as positive or negative (i.e. active/inactive; offence/defence).
However, the binary classification changes as parameters $V_{min}$ and $T_{vel}$ change.
Therefore, we measure the performance of our procedure by evaluating the AUC with respect to different values of $V_{min}$ and $T_{vel}$ used as thresholds.

The proposed strategy is adopted for identifying the best parameters using active/inactive classification.
The same strategy could be applied, making appropriate adaptations, using offence/defence classification. 
Let $\tilde{\textbf{X}}$ be the set of measurements obtained from \textbf{X} by aggregating the observations at a frequency of $1$ second. 
We let $Y\tilde{t}$ be the variable assuming value $1$ if, according to the report, the game is inactive at second $\tilde{t}$, $0$ otherwise.
Moreover, for a given $V_{min}$ and $T_{vel}$ combination, let $Y^\star\tilde{t}$ be the variable assuming value 1 in $\tilde{t}$ if the majority of the observations corresponding to that $\tilde{t}$ was labelled as inactive by the procedure, $0$ otherwise.
We define true positives (TP), true negatives (TN), false positives (FP) and false negatives (FN) accordingly and we compute sensitivity 
%
%
and 1 - specificity.
The method is defined by the following $3$ steps:
\begin{enumerate}
	\item For a given $V_{min}$ we compute the AUC for $T_{vel}$ varying in a range of values. The AUC is then computed for all the $V_{min}$ in a range of values.
	\item $V_{min}$ is selected such that AUC is maximized.
	\item Adopting the Youden's index criteria (Youden's~\cite{youden50index}, Fluss et al.~\cite{fluss2005estimation} and Liu~\cite{liu2012accuracy}), for the chosen $V_{min}$, the value of $T_{vel}$ is selected such that the sum of sensitivity and specificity is maximized.
\end{enumerate} 
\section{Results}
\label{sec:resu}
\begin{figure}[!htb]
	\begin{minipage}{0.5\textwidth}
		\centering
		\includegraphics[width=0.75\linewidth]{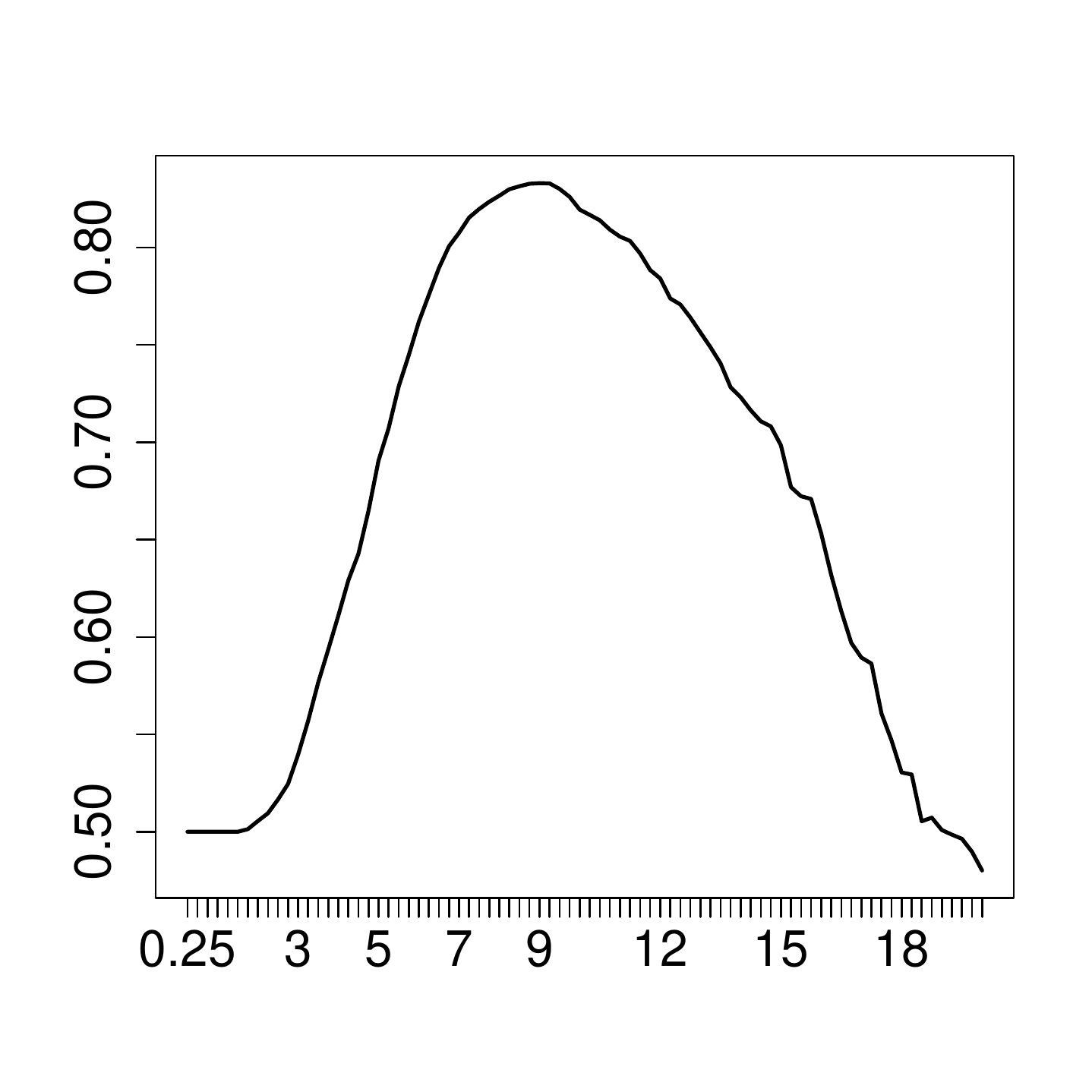}
		\caption{$AUC$ (y-axis) for $V_{min}$ in [0,20] (x-axis).\label{fig:auc_vmin}}
	\end{minipage} \hfill
	\begin {minipage}{0.5\textwidth}
	\centering
	\includegraphics[width=0.75\linewidth]{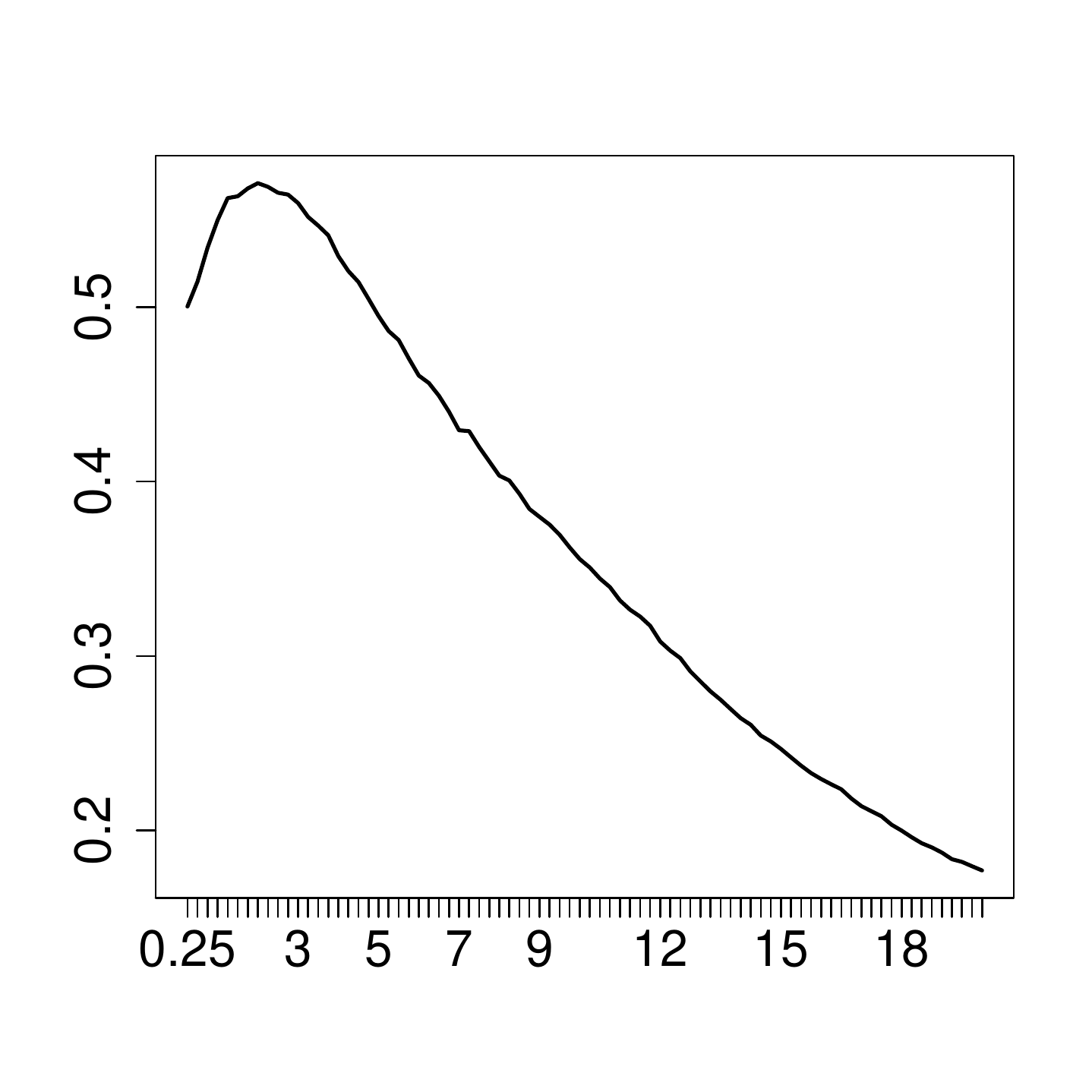}
	\caption{Youden (y-axis) for $T_{vel}$ in [0,20] (x-axis).\label{fig:youden_tvel}}
\end{minipage}
\end{figure}

We apply the method for classifying active/inactive periods to the set of measurements of one game played by a team during the Italian Basketball Cup Final Eight 2017.
The game will be indicated as case study 1 (CS1) and the corresponding set, counting for $505,291$ measurements, will be denoted with $\textbf{X}_1$.
We first compute the value of the $AUC$ corresponding to different values of $V_{min}$.
Up to a given point, the $AUC$ increases as $V_{min}$ increases; beyond such a point, the $AUC$ decreases as $V_{min}$ increases (Figure~\ref{fig:auc_vmin}).
The largest value of $AUC$ is 0.8329.
According to the second step, we select the value of $V_{min}$ corresponding to the largest $AUC$, which is equal to to $9.25 km/h$.
Moving to the third step, we search for the value of $T_{vel}$ that maximize the Youden's index (Figure~\ref{fig:youden_tvel}).
The largest Youden's index is found for $T_{vel} = 2$.
\section{Conclusions}
\label{sec:conc}
%
%
In this work we outline a methodological approach that should be used to automatically select the correct portion of spatial tracking data of a game by just selecting those that correspond to active moments, and to correctly classify them by type of possession. The procedure, along with the identified values for the kinematic parameters may helps experts and analysts who want to analyse tracked data without watching the video of the game.
Future work will focus on a more extensive application of this methodological approach and to a larger number of real case studies.
First by focusing on finding the parameters that best classify active and inactive moments, then by looking to the best choice with regards to the classification among offensive and defensive possessions.
\begin{acknowledgement}
Research carried out in collaboration with the Big\&Open Data Innovation Laboratory (BODaI-Lab), University of Brescia (project nr. 03-2016, title \textit{Big Data Analytics in Sports}, \url{https://bdsports.unibs.it/}), granted by Fondazione Cariplo and Regione Lombardia.
\end{acknowledgement}
%
%
%

\end{document}